\documentstyle[preprint,aps]{revtex}
\begin{document}
\draft
\tightenlines
\title{Renormalization Group and $1/N$ expansion for the three-dimensional Ginzburg-Landau-Wilson model}
\author{Ian D. Lawrie and Dominic J. Lee\thanks{Present address:  Department of Physics,
Simon Fraser University, Burnaby, BC, Canada}}
\address{Department of Physics and Astronomy, University of Leeds, Leeds LS2 9JT, England}
\maketitle
\begin{abstract}
A renormalization-group scheme is developed for the 3-dimensional O($2N$)-symmetric
Ginzburg-Landau-Wilson model, which is consistent with the use of a $1/N$ expansion as a
systematic method of approximation. It is motivated by an application to the critical properties of
superconductors, reported in a separate paper.  Within this scheme, the infrared stable fixed point
controlling critical behaviour appears at $z=0$, where $z=\lambda^{-1}$ is the inverse of the quartic
coupling constant, and an efficient renormalization procedure consists in the minimal subtraction
of ultraviolet divergences at $z=0$.  This scheme is implemented at next-to-leading order, and the
standard results for critical exponents calculated by other means are recovered. An apparently
novel result of this non-perturbative method of approximation is that corrections to scaling (or
confluent singularities) do not, as in perturbative analyses, appear as simple power series in the variable
$y=zt^{\omega\nu}$.  At least in three dimensions, the power series are modified by powers of
$\ln y$.
\end{abstract}
\section{Introduction}

Renormalization-group techniques have long been established as the standard formal framework
for understanding critical phenomena \cite{fisher74,domb,zinn,fisher98}. Those based on
effective Hamiltonians of the Ginzburg-Landau-Wilson type require some kind of supplementary
approximation if concrete information is to be extracted from them, and this is normally provided
by perturbation theory, in the form either of the $\epsilon$ expansion introduced originally by
Wilson and Fisher \cite{wilson} or of a fixed-dimension calculation \cite{parisi}.  For $O(N)$
symmetric models, a systematic expansion in powers of $1/N$ provides an alternative
method of approximation \cite{abe72,ma,kondor,sachdev}.  While the methods that have been used
to extract estimates of critical exponents as power series in $1/N$ are informed by
renormalization-group ideas, there seems to exist no method for implementing
this approximation scheme within a systematic renormalization-group analysis.  The purpose
of this paper is to describe a renormalization scheme which does this.

As is well known, the few terms of the $1/N$ expansions for critical exponents that it has
been possible to calculate explicitly do not yield accurate results for the small values of $N$ that
characterize physically interesting systems, so the development of a new formal technique is
not especially useful from this point of view.  Our motivation for developing this scheme
arises from our efforts to estimate critical-point scaling functions for thermodynamic properties
of superconductors in applied magnetic fields, in particular the specific heat, for which experimental
data have been available for some time (see, for example \cite{overend}).  For this problem, it turns
out that perturbation theory does not yield a well-controlled sequence of approximations.  As
reported in an accompanying paper \cite{lee}, the $1/N$ expansion is advantageous in this respect,
and yields qualitatively encouraging results, although quantitative agreement with existing data
is still hard to achieve.

In outline, the renormalization-group strategy we propose works as follows.  Consider, for example, the
order-parameter susceptibility $\chi(t_0,\lambda_0,\Lambda)$, where $t_0$ is proportional to $T-T_{\mathrm{c}}$,
$\lambda_0$ is a coupling constant which can be taken as temperature-independent in the critical
region and $\Lambda$ is a momentum cutoff of the order of the inverse of a typical interatomic distance.
In the limit $t\to 0$, we expect this susceptibility to diverge as $\chi\sim t_0^{-\gamma}$, where $\gamma$
is a universal exponent which we would like to determine.  In general, a renormalization scheme introduces a
renormalized susceptibility $\chi_{\mathrm{R}}$, together with renormalized parameters $t$ and $\lambda$, which
depend on an arbitrary length scale, say $\mu^{-1}$.  The fact that physics is independent of $\mu$ leads to a
relation of the form
\begin{equation}
\chi_{\mathrm{R}}(t,\lambda,\Lambda/\mu)=P_\chi(\ell)\chi_{\mathrm{R}}(t(\ell),\lambda(\ell),\Lambda/(\ell\mu))
\label{rgchi}
\end{equation}
where $\ell$ is an arbitrary number.  The prefactor $P_\chi(\ell)$ and the running variables $t(\ell)$ and
$\lambda(\ell)$ depend on how the renormalization has been implemented.  Now, the function
$\chi_{\mathrm{R}}(t,\lambda,\Lambda/\mu)$ diverges when its first argument, $t$, approaches zero, but the nature
of this singularity is hard to determine directly.  Suppose, however, that the renormalization scheme has been
constructed in such a way that
\begin{equation}
P_\chi(\ell)\sim\ell^{-\gamma^*/\nu^*}\qquad t(\ell)\sim t\ell^{-1/\nu^*}\qquad \lambda(\ell)\to\lambda^*
\end{equation}
in the limit $\ell\to 0$, where $\gamma^*$, $\nu^*$ and $\lambda^*$ (the `fixed-point' coupling) are
constants. By setting $\ell=t^{\nu^*}$, we find in the limit $t\to 0$
\begin{equation}
\chi_{\mathrm{R}}(t,\lambda,\Lambda/\mu)\sim t^{-\gamma^*}\chi_{\mathrm{R}}(1,\lambda^*,\Lambda/(t^{\nu^*}\mu)).
\label{limchi}
\end{equation}
The exponent $\gamma^*$ will be equal to the true critical exponent $\gamma$
if $\chi_{\mathrm{R}}(1,\lambda^*,\infty)$ has a finite, nonzero value and the renormalization scheme must be
designed to ensure that this is so.  The exponent $\nu^*$ will then be equal to the critical exponent $\nu$,
which characterizes the divergence of the correlation length.

These calculations cannot be carried out exactly, so a renormalization scheme must be supplemented by some
systematic means of approximation and its details will depend on this method of approximation.  Perturbative
methods of approximation, involving an expansion in powers of $\lambda$, rely on the possibility of expressing
exponents such as $\gamma$ as power series in $\epsilon=4-d$, where $d$ is the spatial dimensionality (although
in practice it proves possible also to implement them directly in three dimensions).  Near four dimensions, the
bare susceptibility $\chi(t_0,\lambda_0,\Lambda)$ and other thermodynamic functions diverge as $\Lambda\to\infty$,
so the principal requirement of a renormalization scheme is to remove these divergences from renormalized
functions such as $\chi_{\mathrm{R}}$. One then finds that $\lambda^*$ is of order $\epsilon$, so the perturbative
analysis can be implemented self-consistently.

For the reasons outlined above, we wish to make use of an alternative approximation scheme consisting of an
expansion in powers of $1/N$ with $d=3$.  A systematic means of obtaining the $1/N$ expansion is reviewed in
section II. It is facilitated by an integral transformation of the Hubbard-Stratonovich type which
expresses the partition function as a functional integral over an auxiliary field $\Psi$, with an
effective Hamiltonian $H_{\mathrm{eff}}(\Psi)$.  This partition function can be calculated as a power series
in $1/N$ by the method of steepest descent.  Within this expansion, thermodynamic quantities are most naturally
expressed as functions not of the temperature variable $t$, but rather of the exact inverse susceptibility
$\tilde{t}_0=\chi^{-1}$. The temperature dependence of $\tilde{t}_0$ is given implicitly by solving a constraint
equation of the form $t_0=\Phi(\tilde{t}_0,\lambda_0, \Lambda)$.  At lowest order, this constraint equation simply
locates the saddle point $\Psi=-i(\tilde{t}_0-t_0)$ of the effective Hamiltonian $H_{\mathrm{eff}}(\Psi)$.

In section III, a renormalization scheme is exhibited which manifestly extracts the correct critical exponents
at leading order. In the case of the susceptibility, the renormalized constraint equation
$t=\Phi_{\mathrm{R}}(\tilde{t},\lambda,\Lambda/\mu)=\Phi_{\mathrm{R}}(\chi_{\mathrm{R}}^{-1},\lambda,\Lambda/\mu)$
obeys a relation equivalent to (\ref{rgchi}), but with the roles of $t$ and $\chi_{\mathrm{R}}^{-1}$ reversed.
The substantive issue, addressed in section IV, is how this renormalization scheme can properly be extended to
higher orders.  The usual perturbative strategy fails at this point, because in 3 dimensions the constraint function
$\Phi$ and other thermodynamic quantities remain finite as $\Lambda\to\infty$.  For this reason, we shall
actually omit the cutoff altogether in subsequent sections.  We find, however, that the fixed-point coupling
strength $\lambda^*$ is infinite, and that divergences appear in the limit $\lambda\to\infty$.  For this
reason, we find it convenient to deal with the inverse of the renormalized coupling, $z=\lambda^{-1}$, so that
the fixed point appears at $z^*=0$.  In the context of the $1/N$ expansion, then, the primary requirement
of a renormalization scheme is not to remove divergences that appear as $\Lambda\to\infty$, but rather to
remove those that appear as $z\to 0$.  This crucial observation is the first result that we wish to report.
The second is a concrete renormalization scheme that actually does remove the divergences at $z=0$.  As shown in
section IV, this can be achieved economically by means of a `minimal subtraction' scheme which subtracts
powers of $\ln z$.

We have, of course, checked that the critical exponents obtained from our renormalization scheme
agree with those obtained long ago by other methods.  However, our purpose in devising this novel
renormalization-group strategy is not simply to reproduce old results by a different method.  The application
of this formalism that we have mainly in view is the calculation of scaling functions.  In particular,
the specific heat of a superconductor near its critical point, and in the presence of a magnetic field
$B$, is expected to assume the scaling form $C(T,B)\approx B^{-\alpha/2\nu}{\cal C}(tB^{-1/2\nu})$ and we wish to
estimate the scaling function ${\cal C}(x)$, for which experimental data are available.  When the
Ginzburg-Landau model is enlarged to include the magnetic field, the renormalized specific heat obeys
a relation similar to (\ref{rgchi}), namely
\begin{equation}
C_{\mathrm{R}}(\tilde{t},z,B)=P_C(\ell)C_{\mathrm{R}}(\tilde{t}(\ell),z(\ell),B\ell^{-2})
\stackrel{\ell\to 0}{\longrightarrow}\ell^{-\alpha/\nu}C_{\mathrm{R}}(\ell^{-(2-\eta)}\tilde{t},0,\ell^{-2}B).
\end{equation}
By setting $\tilde{t}(\ell)=1$ and $\ell=B^{1/2}L$, we can identify the scaling function as ${\cal C}(x)
=L^{-\alpha/2\nu}C_{\mathrm{R}}(1,0,L^{-2})$, with $L$ determined as a function of $x$ by solution of the
constraint equation, which then has the form $x=L^{1/\nu}\Phi(1,0,L^{-2})$.  Details of this calculation
are presented in \cite{lee}; here, our principal concern is to explain the renormalization-group formalism
that makes it feasible.  It happens that special considerations apply to the critical behaviour of the
specific heat \cite{abe} and the impact of these on the renormalization-group formalism is discussed in
section V.

While the main purpose of this paper is to expose details of the renormalization scheme that
will be applied in \cite{lee}, we also wish to report what we believe to be a novel feature of the
renormalization group, revealed by the non-perturbative nature of the $1/N$ expansion, that is of some
theoretical interest.  The leading corrections to asymptotic critical singularities (sometimes described
as confluent singularities) appear in the form of a scaling variable
$y=(\lambda-\lambda^*)\vert T-T_{\mathrm{c}}\vert^{\omega\nu}$.
According to conventional wisdom, the order-parameter susceptibility, say, has a singular part
which can be expressed as $\chi^{\mathrm{sing}}=\vert T-T_{\mathrm{c}}\vert^{-\gamma}X(y)$,
where the scaling function $X(y)$ has a power series expansion in $y$.  In perturbation theory,
this is automatically true, but the $1/N$ expansion suggests otherwise.  In 3 dimensions, at
least, we find that $X(y)$ contains, in addition to powers of $y$, singular terms which at next-to-leading
order are of the form $y^n\ln y$.  This is shown in section IV and further discussed, along with our other
principal findings in section VI.

\section{The $1/N$ expansion}
 We consider the standard Ginzburg-Landau-Wilson  theory for $N$ complex fields, defined by
the Hamiltonian density
\begin{equation}
 {\cal H} = \sum^N_{i=1} \left\vert \nabla \phi_i({\bf r}) \right\vert^2+t_0 \left\vert
 \phi_i( {\bf r}) \right\vert^2 + {\lambda_0 \over 4N} \left( \sum^N_{i=1}
 \left\vert \phi_i( {\bf r} ) \right\vert^2 \right)^2\,,
 \end{equation}
in which $t_0$ is taken to be linear in temperature ($t_0\propto T-T_0$, where $T_0$ is the
mean-field transition temperature) and the coupling strength $\lambda_0$ to be
temperature-independent.  This is actually an O(2$N$)-symmetric model; we have assembled its 2$N$ real
fields into $N$ complex fields merely to facilitate the application to a Ginzburg-Landau superconductor
which we have ultimately in view. Introducing sources $j_i(\bbox{r})$ and $j_i^*(\bbox{r})$ for the
fields $\phi_i(\bbox{r})$ and $\phi_i^*(\bbox{r})$, we arrive at the generating functional for
correlation functions
\begin{equation}
Z[j^{\vphantom{*}}_i,j^*_i]=\prod^N_{i=1} \int D \phi_i \int D \phi^*_i \exp\left\{ \int
d^3r \left[-{\cal H}+\sum^N_{i=1} \left(j_i (\bbox{r}) \phi^*_i (\bbox{r})
+j_i^*(\bbox{r}) \phi_i(\bbox{r})\vphantom{\sum}\right) \right]\right\}\,.
\end{equation}
The $1/N$ expansion is generated by the procedure described, for example,
by Br\'ezin {\it et. al.}\cite{brezin}.  A Hubbard-Stratonovich transformation
\begin{equation}
\exp \left[ -\int d^3 r {\lambda_0 \over 4N} \left(
\sum^N_{i=1} | \phi_i |^2 \right)^2 \right] ={\cal N} \int D \Psi
\exp \left[ -\int d^3 r {N \over \lambda_0} \Psi^2 - i \Psi
\sum^N_{i=1} | \phi_i |^2 \right]\,,
\end{equation}
where ${\cal N}$ is a normalization factor whose value is of no consequence,
brings the integral over $\phi_i$ and $\phi^*_i$ to a Gaussian form.  On computing
this integral, we obtain
\begin{equation}
Z[j^{\vphantom{*}}_i,j^*_i]= {\cal N} \int D \Psi \exp \left[ -NH_{\mathrm eff}(\Psi)+ \int d^3r
\int d^3 r' \sum^N_{i=1} j_i({\bf r}) \Delta( \bbox{r},\bbox{r}';\Psi) j^*_i(\bbox{r}') \right],
\end{equation}
where the effective Hamiltonian is
\begin{equation}
H_{\mathrm eff} (\Psi)= \int d^3 r {1 \over \lambda_0} \Psi^2 (\bbox{r}) -
{\mathrm Tr }_{\bbox{r},\bbox{r}'} \ln \Delta (\bbox{r},\bbox{r}';\Psi)
\end{equation}
and the propagator $\Delta (\bbox{r},\bbox{r}';\Psi)$ is the solution of
\begin{equation}
\left[- \nabla^2 +t_0+ i \Psi(\bbox{r})\right]\Delta(\bbox{r},\bbox{r}';\Psi) = \delta(\bbox{r} -\bbox{r}')\,.
\end{equation}
The connected correlation functions $G_{i_1\cdots i_n}(\bbox{r}_1,\cdots\bbox{r}_n)
=\langle\phi_{i_1}^*(\bbox{r}_1)\cdots\phi_{i_n}(\bbox{r}_n)\rangle_{\mathrm c}$ are, of course,
obtained by differentiation of $Z[j_i,j_i^*]$.  For example, the two-point function
$G_{ij}(\bbox{r},\bbox{r}')\equiv\delta_{ij}G^{(2)}(\bbox{r}-\bbox{r}')$ is given by
\begin{equation}
G^{(2)}(\bbox{r}_1-\bbox{r}_2)=\langle\phi^*_1(\bbox{r}_1)\phi_1(\bbox{r}_2)\rangle_{\mathrm c}
=\left.\frac{\delta^2\ln Z}{\delta j_1(\bbox{r}_1)\delta j_1^*(\bbox{r}_2)}\right\vert_{j=j^*=0}
=\frac{\int D\Psi\,\Delta(\bbox{r}_1,\bbox{r}_2;\Psi)e^{-NH_{\mathrm eff}(\Psi)}}
{\int D\Psi\,e^{-NH_{\mathrm eff}(\Psi)}}\,.
\end{equation}
Similarly, we can define a four-point function
\begin{equation}
G^{(4)}(\bbox{r}_1,\bbox{r}_2;\bbox{r}_3,\bbox{r}_4)
=\frac{\int D\Psi\,\Delta(\bbox{r}_1,\bbox{r}_2;\Psi)\Delta(\bbox{r}_3,\bbox{r}_4;\Psi)e^{-NH_{\mathrm eff}(\Psi)}}
{\int D\Psi\,e^{-NH_{\mathrm eff}(\Psi)}}\,.
\end{equation}

Owing to the factor of $N$ that multiplies $H_{\mathrm eff}(\Psi)$, the $1/N$ expansion is now generated by
the method of steepest descent.  It proves convenient to formulate this expansion in terms of a parameter
$\tilde{t}_0$, defined by
\begin{equation}
\tilde{t}_0=\Gamma^{(2)}(0)\,,
\label{ttildedef}
\end{equation}
where $\Gamma^{(2)}(\bbox{p})$ is the inverse of the Fourier transform of $G^{(2)}(\bbox{r}-\bbox{r}')$.
We write
\begin{equation}
\Psi(\bbox{r}) = -i \left(\tilde{t}_0-t_0-\frac{1}{N}\delta\right) + (2N)^{-1/2} \psi(\bbox{r})\,,
\end{equation}
where $\delta$ is defined by imposing the constraint $\langle\psi(\bbox{r})\rangle=0$.  Since $\delta$ has
contributions of all orders in $1/N$, we set $\delta = \delta_0+N^{-1}\delta_1+\cdots$.  The position-independent
value $\Psi_0=-i(\tilde{t}_0-t_0)$ locates the saddle point of $H_{\mathrm eff}$ up to corrections of order $1/N$,
and the propagator $\Delta(\bbox{r},\bbox{r}';\Psi)$ is given by
$\Delta(\bbox{r},\bbox{r}';\Psi)=\Delta(\bbox{r}-\bbox{r}')+O(N^{-1/2})$, where
\begin{equation}
\Delta(\bbox{r}) =\int\frac{d^3p}{(2\pi)^3}\frac{e^{i\bbox{p}.\bbox{r}}}{(p^2+\tilde{t}_0)}\,.
\end{equation}
The two quantities $NH_{\mathrm eff}(\Psi)$ and $\Delta(\bbox{r},\bbox{r}';\Psi)$ can now be expanded
in powers of $1/N$.  For $NH_{\mathrm eff}(\Psi)$, discarding irrelevant terms that are independent of
$\psi(\bbox{r})$, the result is
\begin{eqnarray}
NH_{\mathrm eff}&=&iN^{1/2}\frac{\sqrt{2}}{\lambda_0}\left[t_0-f(\tilde{t}_0,\lambda_0)\right]\int d^3r\,\psi(\bbox{r})
\nonumber\\
&&+\frac{1}{2}\int d^3r\,d^3r'\,\psi(\bbox{r})D^{-1}(\bbox{r}-\bbox{r'})\psi(\bbox{r}')\nonumber\\
&&-iN^{-1/2}\frac{1}{6\sqrt{2}}\int d^3r\,d^3r'\,d^3r''\,\Delta_3(\bbox{r},\bbox{r}',\bbox{r}'')\psi(\bbox{r})\psi(\bbox{r}')\psi(\bbox{r}'')
\nonumber\\
&&-N^{-1}\left[\frac{1}{16}\int d^3r\,d^3r'\,d^3r''\,d^3r'''\,\Delta_4(\bbox{r},\bbox{r}',\bbox{r}'',\bbox{r}''')
\psi(\bbox{r})\psi(\bbox{r}')\psi(\bbox{r}'')\psi(\bbox{r}''')\right.\nonumber\\
&&\qquad\qquad\left.-\frac{1}{2}\delta_0\int d^3r\,d^3r'\,d^3r''\,\Delta_3(\bbox{r},\bbox{r}',\bbox{r}'')\psi(\bbox{r})\psi(\bbox{r}')\right]
+O\left(N^{-3/2}\right)\,,
\label{hexpansion}
\end{eqnarray}
where the $\psi$ propagator is given by
\begin{equation}
D^{-1}(\bbox{r}-\bbox{r}') = \lambda_0^{-1}\delta(\bbox{r}-\bbox{r}') + \frac{1}{2} \Delta(\bbox{r}-\bbox{r}')
\Delta(\bbox{r}'-\bbox{r})
\end{equation}
and the vertex functions are
\begin{equation}
\Delta_n(\bbox{r}_1,\cdots,\bbox{r}_n)=\Delta(\bbox{r}_1-\bbox{r}_2)\Delta(\bbox{r}_2-\bbox{r}_3)\cdots
\Delta(\bbox{r}_n-\bbox{r}_1)\,.
\end{equation}
The function $f(\tilde{t}_0,\lambda_0)$ is
\begin{equation}
f(\tilde{t}_0,\lambda_0)=\tilde{t}_0-\frac{1}{N}\delta_0-\frac{1}{2}\lambda_0\Delta(\bbox{0})
-\frac{1}{2}\lambda_0\delta_0\int d^3r\,\Delta_2(\bbox{r})\,.
\end{equation}
Although in principle the counterterm $\delta$ is determined by the constraint $\langle\psi(\bbox{r})\rangle=0$
while the definition (\ref{ttildedef}) determines $\tilde{t}_0$ as a function of $t_0$ and $\lambda_0$, we see
from the expansion (\ref{hexpansion}) that practical calculations invert this logic.  Thus, the term proportional
to $N^{-1}\delta_0$ is quadratic in $\psi(\bbox{r})$, so we arrange for (\ref{ttildedef}) to be true by adjusting
the value of $\delta_0$.  On the other hand, the term that is linear in $\psi(\bbox{r})$ provides a counterterm
that is adjusted to make $\langle\psi(\bbox{r})\rangle$ vanish, yielding a constraint (or gap) equation
\begin{equation}
t_0=f(\tilde{t}_0,\lambda_0)+O\left(N^{-1}\right)
\end{equation}
that implicitly determines $\tilde{t}_0$ as a function of $t_0$ and $\lambda_0$.

For the propagator $\Delta(\bbox{r},\bbox{r}';\Psi)$ we have the expansion
\begin{eqnarray}
\Delta(\bbox{r},\bbox{r}';\Psi)&=&\Delta(\bbox{r}-\bbox{r}')
-iN^{-1/2}\frac{1}{\sqrt{2}}\int d^3r_1\,\bar{\Delta}_2(\bbox{r},\bbox{r}';\bbox{r}_1)\psi(\bbox{r}_1)\nonumber\\
&&-N^{-1}\left[\frac{1}{2}\int d^3r_1\,d^3r_2\,\bar{\Delta}_3(\bbox{r},\bbox{r}';\bbox{r}_1,\bbox{r}_2)
\psi(\bbox{r}_1)\psi(\bbox{r}_2)\right.\nonumber\\
&&\qquad\qquad\left.+\delta_0\int d^3r_1\,\bar{\Delta}_2(\bbox{r},\bbox{r}';\bbox{r}_1)\right]\nonumber\\
&&+iN^{-3/2}\left[\frac{1}{2\sqrt{2}}\int d^3r_1\,d^3r_2\,d^3r_3\,\bar{\Delta}_4(\bbox{r},\bbox{r}';\bbox{r}_1,
\bbox{r}_2,\bbox{r}_3)\psi(\bbox{r}_1)\psi(\bbox{r}_2)\psi(\bbox{r}_3)\right.\nonumber\\
&&\qquad\qquad\left.+\sqrt{2}\delta_0\int d^3r_1\,d^3r_2\,\bar{\Delta}_3(\bbox{r},\bbox{r}';\bbox{r}_1,\bbox{r}_2)
\psi(\bbox{r}_1)\right]+O(N^{-2})\,,
\label{deltaexpansion}
\end{eqnarray}
with
\begin{equation}
\bar{\Delta}_n(\bbox{r},\bbox{r}';\bbox{r}_1,\cdots,\bbox{r}_{n-1})=\Delta(\bbox{r}-\bbox{r}_1)
\Delta(\bbox{r}_1-\bbox{r}_2)\cdots\Delta(\bbox{r}_{n-1}-\bbox{r}')\,.
\end{equation}

The two expansions given in (\ref{hexpansion}) and (\ref{deltaexpansion}) can be summarized
diagrammatically by the Feynman rules given schematically in figure 1 for calculating the correlation
functions $G_{i_1\cdots i_n}(\bbox{r}_1,\cdots\bbox{r}_n)$.  In practice, it is most convenient to
interpret these in momentum space.  Then the $\phi$ and $\psi$ propagators are
\begin{eqnarray}
\Delta(\bbox{p})&=&\left(p^2+\tilde{t}_0\right)^{-1}\\
D(\bbox{p})&=&\left[\lambda_0^{-1}+\Pi(\bbox{p})\right]^{-1}\,,
\end{eqnarray}
with
\begin{equation}
\Pi(\bbox{p})=\frac{1}{2}\int\frac{d^3k}{(2\pi)^3}\frac{1}{[\bbox{k}^2+\tilde{t}_0][(\bbox{k}+\bbox{p})^2+\tilde{t}_0]}
=\frac{1}{8\pi\vert\bbox{p}\vert}\tan^{-1}\left(\frac{\vert\bbox{p}\vert}{2\tilde{t}_0^{1/2}}\right)\,.
\end{equation}
For future use, we note that $\Pi(0)=a\tilde{t}_0^{-1/2}$, with $a=1/16\pi$, while for large momenta we can write
\begin{equation}
\Pi(\bbox{p})=\frac{b}{\vert\bbox{p}\vert}-\frac{\tilde{t}_0^{1/2}}{p^2}\widetilde{\Pi}\left(\frac{\tilde{t}_0}{p^2}\right)\,,
\end{equation}
with $b=1/16$ and
\begin{equation}
\widetilde{\Pi}(\tau)=\frac{1}{8\pi}\tau^{-1/2}\tan^{-1}(2\tau^{1/2})=4a\left(1-\frac{4}{3}\,\tau +\cdots\right)\,.
\end{equation}

For our immediate purposes, we require explicit expressions at next-to-leading order in $1/N$ for the constraint
equation $\langle\psi(\bbox{r})\rangle=0$, the two-point function $G^{(2)}(\bbox{p})$ and the truncated four-point
function $\Gamma^{(4)}(\bbox{p}_i)=G^{(4)}(\bbox{p}_i)/\prod_{j=1}^4G^{(2)}(\bbox{p}_j)$, which is
one-particle-irreducible with respect to the $\phi$ propagator $\Delta(\bbox{p})$.  These are shown diagrammatically
in figures 2, 3 and 4 respectively.  The constraint equation is
\begin{eqnarray}
t_0&=&f(\tilde{t}_0,\lambda_0)+\frac{\lambda_0}{4N}A+O\left(N^{-2}\right)\nonumber\\
&=&\tilde{t}_0-\frac{\lambda_0}{2}\int\frac{d^3k}{(2\pi)^3}\Delta(\bbox{k})
+N^{-1}\left[\frac{\lambda_0}{4}A(\tilde{t}_0,\lambda_0)-\lambda_0\delta_0\Pi(0)-\delta_0\right]+O\left(N^{-2}\right)\,,
\end{eqnarray}
where $A(\tilde{t}_0,\lambda_0)$ is the integral
\begin{equation}
A(\tilde{t}_0,\lambda_0)=\int\frac{d^3k}{(2\pi)^3}\frac{d^3k'}{(2\pi)^3}\Delta(\bbox{k})^2\Delta(\bbox{k}')
D(\bbox{k}+\bbox{k}')\,.\label{aintegral}
\end{equation}
The inverse two-point function $\Gamma^{(2)}(\bbox{p})=G^{(2)}(\bbox{p})^{-1}$ is given by
\begin{equation}
\Gamma^{(2)}(\bbox{p})=p^2+\tilde{t}_0+N^{-1}\left[\Sigma(\bbox{p};\tilde{t}_0,\lambda_0)-\delta_0\right]+O\left(N^{-2}\right)\,,
\end{equation}
with
\begin{equation}
\Sigma(\bbox{p};\tilde{t}_0,\lambda_0)=\frac{1}{2}\int\frac{d^3k}{(2\pi)^3}\Delta(\bbox{k}+\bbox{p})D(\bbox{k})\,.
\end{equation}
Thus, the definition (\ref{ttildedef}) of $\tilde{t}_0$ allows us to identify
\begin{equation}
\delta_0=\Sigma(\bbox{0};\tilde{t}_0,\lambda_0)\,,
\label{deltavalue}
\end{equation}
and the constraint equation is given explicitly by
\begin{equation}
t_0=\tilde{t}_0-\frac{\lambda_0}{2}\int\frac{d^3k}{(2\pi)^3}\Delta(\bbox{k})
+N^{-1}\left[\frac{\lambda_0}{4}A(\tilde{t}_0,\lambda_0)-
\Sigma(\bbox{0};\tilde{t}_0,\lambda_0)\left[1+\lambda_0\Pi(0)\right]\right]+O\left(N^{-2}\right)\,.
\label{constraint}
\end{equation}
We shall need the four-point function $\Gamma^{(4)}(\bbox{p}_i)$ evaluated at $\bbox{p}_i=0$.
For this case, the multiloop integrals shown in Fig. 4 can be simplified by using the results
\begin{eqnarray}
\int\frac{d^3k'}{(2\pi)^3}\Delta(\bbox{k}')^2\Delta(\bbox{k}+\bbox{k}')&=&-\frac{d}{d\tilde{t}_0}
\Pi(\bbox{k})\nonumber\\
&=&\frac{2a\tilde{t}_0^{-1/2}}{k^2+4\tilde{t}_0}\\
\int\frac{d^3k'}{(2\pi)^3}\left[\Delta(\bbox{k}')^2\Delta(\bbox{k}+\bbox{k}')^2
+2\Delta(\bbox{k}')^3\Delta(\bbox{k}+\bbox{k}')\right]&=&\frac{d^2}{d{\tilde{t}_0}^2}\Pi(\bbox{k})
\nonumber\\
&=&a\tilde{t}_0^{-3/2}\left[\frac{1}{k^2+4\tilde{t}_0}+\frac{8\tilde{t}_0}{(k^2+\tilde{t}_0)^2}\right]
\end{eqnarray}
together with the value of $\delta_0$ given in (\ref{deltavalue}).  We then find
\begin{equation}
\Gamma^{(4)}(\bbox{0})=N^{-1}D(\bbox{0})+N^{-2}\left[B_1(\tilde{t}_0,\lambda_0)
+B_2(\tilde{t}_0,\lambda_0)+B_3(\tilde{t}_0,\lambda_0)\right]+O(N^{-3})\,,
\end{equation}
where the remaining integrals are
\begin{eqnarray}
B_1(\tilde{t}_0,\lambda_0)&=&\frac{1}{4}a\tilde{t}_0^{-1/2}D(0)^2\int\frac{d^3k}{(2\pi)^3}D(\bbox{k})
\left[\frac{8}{(k^2+4\tilde{t}_0^2)^2}-\frac{3}{(k^2+\tilde{t}_0)(k^2+4\tilde{t}_0)}\right]\label{b1integral}\\
B_2(\tilde{t}_0,\lambda_0)&=&-D(0)\int\frac{d^3k}{(2\pi)^3}D(\bbox{k})\frac{1}{(k^2+\tilde{t}_0^2)^2}\label{b2integral}\\
B_3(\tilde{t}_0,\lambda_0)&=&-\int\frac{d^3k}{(2\pi)^3}D(\bbox{k})^2\left[\frac{1}{k^2+\tilde{t}_0}
-a\tilde{t}_0^{-1/2}D(0)\frac{1}{k^2+4\tilde{t}_0}\right]^2\,,\label{b3integral}
\end{eqnarray}
with $D(0)=\left(\lambda_0^{-1}+a\tilde{t}_0^{-1/2}\right)^{-1}$.

\section{Renormalization group at leading order}
In the limit $N\to\infty$, the one-particle-irreducible four-point function $\Gamma^{(4)}(\bbox{p}_i)$
vanishes, and so do the higher multi-point functions.  The theory is therefore effectively Gaussian,
all correlations being determined by the two-point function $G^{(2)}(\bbox{p})=(p^2+\tilde{t}_0)^{-1}$.
The analysis of critical-point behaviour now reduces essentially
to determining the dependence of $\tilde{t}_0$ on the temperature-like variable $t_0$ from the
constraint equation (\ref{constraint}).  This equation can be written as
\begin{equation}
t_0-t_{0{\mathrm c}}=\tilde{t}_0+2a\lambda_0\tilde{t}_0^{1/2}\,,
\label{leadingconstraint}
\end{equation}
where $t_{0{\mathrm c}}$ is the critical value of $t_0$, corresponding to $\tilde{t}_0=0$.  It represents
fluctuation corrections to the mean-field transition temperature and is given formally by the divergent
integral
\begin{equation}
t_{0{\mathrm c}}=-\frac{\lambda_0}{2}\int\frac{d^3k}{(2\pi)^3}\frac{1}{k^2}\,.
\end{equation}
Near the critical temperature, the leading behaviour of $\tilde{t}_0$ is clearly
$\tilde{t}_0\propto (t_0-t_{0{\mathrm c}})^{\gamma}\,$, with $\gamma=2$.  According to standard
renormalization-group ideas, we should be able to find a fixed-point value of the coupling
strength, $\lambda_0^*$, for which this power law is exact, and this value is clearly
$\lambda_0^*=\infty$.  Within a suitable renormalization scheme, we might expect this fixed
point to correspond to a finite value of a renormalized coupling $\lambda$.  The usual motivations
for renormalization, namely the removal of ultraviolet divergences or the exponentiation of
logarithms of $t_0-t_{0{\mathrm c}}$ are absent here, because the constraint equation and
correlation functions are finite, and contain no logarithms.  Nevertheless, let us define
a renormalized temperature variable $t$, a renormalized coupling $\lambda$ and a renormalized
inverse susceptibility $\tilde{t}$ by
\begin{eqnarray}
\lambda_0&=&\mu Z_{\lambda}(\lambda)\lambda\\
t_0&=&t_{0{\mathrm c}}+\mu^2 Z_t(\lambda)t\\
\tilde{t}_0&=&\mu^2\tilde{t}
\end{eqnarray}
where, as usual, $\mu$ is an arbitrary renormalization scale which serves to make
$t$, $\tilde{t}$ and $\lambda$ dimensionless.  We specify the renormalization factors $Z_t$ and $Z_\lambda$
by the normalization conditions
\begin{eqnarray}
\Gamma^{(2)}(p^2=0,t=1)&=&\mu^2\,,\label{nc1}\\
\lim_{N\to\infty}N\Gamma^{(4)}(\bbox{p}_i=\bbox{0},t=1)&=&\mu\lambda\,.
\label{nc2}
\end{eqnarray}
In practice, our Feynman rules yield correlation functions as functions of $\tilde{t}$ rather than $t$,
and these conditions are more conveniently expressed as the requirements that
$t_0-t_{0{\mathrm c}}=Z_t\mu^2$ and $D(0)=\mu\lambda$ when $\tilde{t}=1$.  We find
\begin{eqnarray}
Z_{\lambda}(\lambda)&=&(1-a\lambda)^{-1}\\
Z_t(\lambda)&=&\frac{1+a\lambda}{1-a\lambda}\,.
\end{eqnarray}

Expressed in terms of $\lambda$ and $t$, the constraint equation (\ref{leadingconstraint}) is
\begin{equation}
t=(1+a\lambda)^{-1}\left[(1-a\lambda)\tilde{t}+2a\lambda\tilde{t}^{1/2}\right]\,.
\label{renconstraint}
\end{equation}
Clearly, on setting $\lambda=\lambda^*=a^{-1}$, we do find the exact power-law behaviour
$t=\tilde{t}^{1/\gamma}$.  More generally, using the scaling fields $z=\lambda^{-1}-a$
and $\tau=(1+z/2a)t$, we can express the inverse susceptibility $\tilde{t}$ in the scaling form
\begin{equation}
\tilde{t}=\tau^{\gamma}{\cal T}(x)\,,
\label{scalingconstraint}
\end{equation}
where the scaling variable is $x=z\tau^{\omega\nu}/2a$ with $\omega\nu=1$ and the universal
scaling function is given by
\begin{equation}
{\cal T}(x)=(2x)^{-2}\left[1+2x-\sqrt{1+4x}\,\right]=1-2x+5x^2+O(x^3)\,.
\label{scalingfunction}
\end{equation}
Formally, the scaling relation (\ref{scalingconstraint}) and the values of the exponents can be
deduced from the renormalization-group equation, which we find it convenient to formulate
in terms of the variables $\tilde{t}$ and $z$.  Typically, given a thermodynamic quantity
$A(\lambda_0,\tilde{t}_0)$, its renormalized counterpart $A_{\mathrm R}(z,\tilde{t})$ is given by
\begin{equation}
A(\lambda_0,\tilde{t}_0)=Z_A(z)\mu^{D_A}A_{\mathrm R}(z,\tilde{t})\,,
\end{equation}
where $D_A$ is the canonical dimension of $A$.  The criterion for determining the
renormalization factor $Z_A(z)$ will be discussed in the next section.  The fact that
$A(\lambda_0,\tilde{t}_0)$ is independent of the renormalization scale $\mu$ is expressed by the
renormalization-group equation
\begin{equation}
\left[\beta(z)\frac{\partial}{\partial z}-\left(\vphantom{\sum}2-\eta(z)\right)\tilde{t}\frac{\partial}{\partial \tilde{t}}
+D_A+\sigma_A(z)\right]A_{\mathrm R}(z,\tilde{t})=0\,,
\label{rgeqn}
\end{equation}
where
\begin{eqnarray}
\beta(z)&=&\mu\left(\frac{\partial z}{\partial\mu}\right)_{\lambda_0,\,\tilde{t}_0}\\
2-\eta(z)&=&-\frac{\mu}{\tilde{t}}\left(\frac{\partial \tilde{t}}{\partial\mu}\right)_{\lambda_0,\,\tilde{t}_0}\\
\sigma_A(z)&=&\beta(z)\frac{d\ln Z_A(z)}{dz}\,.
\end{eqnarray}
Solution of this equation by the method of characteristics yields the relation
\begin{equation}
A_{\mathrm R}(z,\tilde{t})=P_A(\ell)A_{\mathrm R}\left(\vphantom{\sum}z(\ell),\tilde{t}(\ell)\right)\,,
\label{rgsolution}
\end{equation}
where $\ell$ is an arbitrary number and $z(\ell)$ is the solution of
\begin{equation}
\ell\frac{dz(\ell)}{d\ell}=\beta\left(\vphantom{\sum}z(\ell)\right)
\end{equation}
with the initial condition $z(1)=z$, while $\tilde{t}(\ell)$ and the prefactor $P_A(\ell)$ are given by
\begin{eqnarray}
\tilde{t}(\ell)&=&\tilde{t}\ell^{-2}\exp\left[\int_1^{\ell}\frac{d\ell'}{\ell'}\eta\left(\vphantom{\sum}z(\ell')\right)\right]\\
P_A(\ell)&=&\exp\left\{\int_1^{\ell}\frac{d\ell'}{\ell'}\left[D_A+\sigma_A\left(\vphantom{\sum}z(\ell')\right)\right]\right\}\,.
\end{eqnarray}
In the case at hand, taking $A=t_0-t_{0{\mathrm c}}$, we find $\beta(z)=z$, $\eta(z)=0$, $D_t=2$ and
$\sigma_t(z)=-2a/(z+2a)$. The characteristic functions are
\begin{equation}
z(\ell)=z\ell\,,\qquad\qquad \tilde{t}(\ell)=\tilde{t}\ell^{-2}\,,\qquad\qquad P_t(\ell)=\ell\left(\frac{z\ell+2a}
{z+2a}\right)\label{characteristicfns}
\end{equation}
and it is easy to verify by substitution into (\ref{renconstraint}) that the relation (\ref{rgsolution}) is
satisfied.  To obtain the scaling form (\ref{scalingconstraint}), we can choose the arbitrary parameter
$\ell$ as the solution of the equation $\tilde{t}(\ell)=1$.  The neighbourhood of the critical point $\tilde{t}=0$
then corresponds (assuming that $\eta(z)<2$) to the limit $\ell\to 0$.  In this limit, we will generally
expect that
\begin{eqnarray}
z({\ell})&\sim&z\ell^{\omega}\\
\tilde{t}(\ell)&\sim&\tilde{t}\ell^{-(2-\eta)}\\
P_t(\ell)&\sim&\ell^{1/\nu}\,,
\end{eqnarray}
where $\omega=\beta'(0)$, $\eta=\eta(0)$ and $\nu=[2+\sigma_t(0)]^{-1}$ are the usual critical exponents.
(Here, of course, we have $\omega=1$, $\eta=0$ and $\nu=1$.) On setting $\ell\sim\tilde{t}^{1/(2-\eta)}$,
the relation (\ref{rgsolution}) becomes
\begin{equation}
t(z,\tilde{t})\sim \tilde{t}^{1/\gamma}t\left(z\tilde{t}^{\omega\nu/\gamma},1\right)\,,
\end{equation}
with $\gamma=(2-\eta)\nu$, which can be inverted to express $\tilde{t}$ in the scaling form
(\ref{scalingconstraint}).

\section{Extension to higher orders}
It is well known that the exponents obtained in the previous section are modified
when $N$ is finite, and can be expressed as power series in $1/N$.  The series for
correlation functions contain infrared singularities, in the form of logarithms of
$\tilde{t}_0$, which must be exponentiated to obtain the correct power laws, and
our aim is to find a renormalization prescription that will effect this exponentiation.
Experience with perturbation theory and the $\epsilon$ expansion suggests that
application of normalization conditions such as (\ref{nc1}) and (\ref{nc2}) should
achieve this, but a strategy of this kind is unsatisfactory for two reasons.  At a purely
practical level, one obtains renormalization-group functions $\beta(z)$, etc. that are
cumbersome (and, indeed, singular) functions of $z$.  More fundamentally, we have, so far,
no reason within the $1/N$ expansion to expect that a renormalization scheme of this kind
really will produce the correct exponentiation.  We should therefore consider just what
kind of renormalization scheme is needed.

A simple criterion becomes apparent, if we assume that the renormalization-group analysis
will produce relations of the kind exhibited in (\ref{rgsolution}).  On the left hand side,
the function $A(z,\tilde{t})$ has infrared singularities when its second argument,
$\tilde{t}$, approaches zero.  On the right hand side, these singularities are removed by
the condition $\tilde{t}(\ell)=1$, but there remains the danger that they might reappear
through the limit $z(\ell)\to 0$.  Evidently, the singularities will be correctly
exponentiated into the prefactor $P_A(\ell)$, provided that
$A\left(\vphantom{\sum}z(\ell),1\right)$ remains finite and non-zero in the limit
$z(\ell)\to 0$.  The primary requirement of a renormalization scheme is to ensure that
this is so.

To see how a renormalization scheme might work, let us examine the divergences
that might occur in the unrenormalized theory when $\tilde{t}_0$ is fixed to a
non-zero value.  The Feynman diagrams generated by the $1/N$ expansion are
topologically similar to those in a theory of two fields $\phi$ and $\psi$, with
propagators $\Delta(\bbox{p})=\left(p^2+\tilde{t}_0\right)^{-1}$
and $D(\bbox{p})=\left[z_0+\Pi(\bbox{p})\right]^{-1}$ respectively, and a single interaction
vertex corresponding to $\phi^2\psi$.  Here, we use the notation $z_0=\lambda_0^{-1}$.
With $\tilde{t}_0$ fixed, there are no infrared divergences.  In the ultraviolet regime
$\vert\bbox{p}\vert\to\infty$, we have $\Delta(\bbox{p})\sim 1/p^2$ and, provided that
$z_0\ne 0$, $D(\bbox{p})\sim {\rm constant}$.  Consider, then, a subintegral that is
one-particle-irreducible with respect to both propagators, that attaches to $m$ external
$\phi$ legs and $n$ external $\psi$ legs, and has $L$ loop integrations.  It is simple to
show that the superficial degree of ultraviolet divergence of this integral is
\begin{equation}
\Omega_{\mathrm uv}(z_0\ne 0)=4-L-m-2n\,.
\end{equation}
There are in fact only two diagrams that diverge, namely those with $L=1$ and $(m,n)=(0,1)$
or $(m,n)=(2,0)$.  These divergences are subtracted by the constraint $\langle\psi\rangle=0$
and by the additive mass renormalization corresponding to $t_{0{\mathrm c}}$.  The remaining
theory is ultraviolet-finite in 3 dimensions.  When $z_0=0$, however, the limiting behaviour of
the $\psi$ propagator is $D(\bbox{p})\sim \vert\bbox{p}\vert$.  In that case, we have
\begin{equation}
\Omega_{\mathrm uv}(z_0= 0)=3-2n-\frac{m}{2}\,.
\end{equation}
There are now additional divergences which will appear as singularities at $z_0=0$, and these
must be removed by further renormalization.  Unfortunately, the freedom that we have to rescale
the original field $\phi$ and the parameters $t_0-t_{0{\mathrm c}}$ and $z_0$ does not yield
a set of independent counterterms  corresponding to the divergent subintegrals.  Indeed, we can
offer no direct proof that this rescaling will suffice to remove all the divergences.  An indirect (and
admittedly heuristic) argument suggesting that all the divergences can nevertheless be removed
is afforded by the observation that the limit $z\to 0$ is equivalent (with $t_{0{\mathrm c}}\sim
-\textrm{constant}\times\lambda_0$) to a limit $\lambda_0\to\infty$, $t_0\to-\infty$, in which the model
studied here should be equivalent to the non-linear $\sigma$ model whose renormalizability to
all orders of the $1/N$ expansion is proved in \cite{arefeva}.

Here, we adopt the following pragmatic approach.  First, we introduce a wavefunction renormalization
factor $Z_{\phi}(z)$, so that correlation functions $\Gamma^{(n)}$ are renormalized according to
\begin{equation}
\Gamma^{(n)}(\bbox{p}_i;\lambda_0,t_0)
=Z_{\phi}(z)^{-n/2}\Gamma^{(n)}_{\mathrm R}(\bbox{p}_i;z,\tilde{t},\mu)\,.
\end{equation}
(These $\Gamma^{(n)}$ are defined as usual by Legendre transformation of the generating functional
$\ln Z[j^{\vphantom{*}}_i,j_i^*]$ and would be one-particle-irreducible when calculated in perturbation theory.)
The definition (\ref{ttildedef}) then implies that $\tilde{t}_0$ is renormalized according to
\begin{equation}
\tilde{t}_0=Z_{\phi}^{-1}(z)\mu^2\tilde{t}
\end{equation}
and we introduce renormalized parameters $z$ and $t$ defined by
\begin{eqnarray}
z_0&=&Z_z(z)\mu^{-1} z\\
t_0&=&t_{0{\mathrm c}}+\mu^2 \frac{(z+2a)}{z}Z_t(z)t\,.\label{trenorm}
\end{eqnarray}
The renormalization factors $Z_{\phi}(z)$, $Z_{z}(z)$ and $Z_{t}(z)$ must be chosen so as to make
the two renormalized correlation functions $\Gamma_{\mathrm R}^{(2)}(\bbox{p}_i=0)$ and
$\Gamma^{(4)}(\bbox{p}=0)$ and the constraint equation for the renormalized temperature variable $t$
finite when $z\to 0$ with $\tilde{t}$ fixed to some nonzero value.  As usual, there are many ways in
which this might be achieved.  Here, we propose a `minimal subtraction' scheme which, however, may
well be susceptible of further refinement for specific purposes.  At relative order $1/N$, we show in
Appendix A that the singular parts of the relevant Feynman diagrams are of the form
$(\ln z)\times(\textrm{infinite power series in }z).$  The minimal way of ensuring finite limits as $z\to 0$
is to subtract just the leading singular terms, proportional to $\ln z$.  Proceeding in this way, we obtain
for the renormalization factors
\begin{eqnarray}
Z_{\phi}(z)&=&1+\frac{1}{N}\frac{1}{6b}S_3\ln z +O(N^{-2})\\
Z_t(z)&=&1-\frac{1}{N}\frac{2}{3b}S_3\ln z+O(N^{-2})\\
Z_z(z)&=&1+\frac{1}{N}\frac{4}{3b}S_3\ln z+O(N^{-2})\,
\end{eqnarray}
and hence for the renormalization-group functions
\begin{eqnarray}
\beta(z)&=&\omega z+O(N^{-2})\label{betaofz}\\
\eta(z)&=&\eta+O(N^{-2})\label{etaofz}\\
2+\sigma_t(z)&=&\frac{1}{\nu}+\frac{\omega z}{z+2a}+O(N^{-2})\,,
\end{eqnarray}
where the exponents are given by
\begin{eqnarray}
\omega&=&1-\frac{1}{N}\frac{4}{3b}S_3+O(N^{-2})=1-\frac{32}{3\pi^2N}+O(N^{-2})\\
\eta&=&\frac{1}{N}\frac{1}{6b}S_3+O(N^{-2})=\frac{4}{3\pi^2N}+O(N^{-2})\\
\nu&=&1-\frac{1}{N}\frac{2}{3b}S_3+O(N^{-2})=1-\frac{16}{3\pi^2N}+O(N^{-2})
\end{eqnarray}
in agreement with standard results (bearing in mind that our $N$ complex fields correspond to $2N$ real fields).
The characteristic functions that were given at leading order by (\ref{characteristicfns}) become
\begin{equation}
z(\ell)=z\ell^{\,\omega}\,,\qquad\qquad \tilde{t}(\ell)=\tilde{t}\ell^{-(2-\eta)}\,,\qquad\qquad
P_t(\ell)=\ell^{1/\nu}\left(\frac{z\ell^{\,\omega}+2a}{z+2a}\right)\,.
\label{newcharacteristics}
\end{equation}

It is at least of formal interest to obtain the scaling form of the inverse susceptibility (\ref{scalingconstraint}) at
next-to-leading order, for the following reason.  At leading order, the scaling function (\ref{scalingfunction}) has
a simple series expansion in powers of $x\sim z\tau^{\omega\nu}$.  In perturbative realizations of the renormalization
group, whether formulated as a systematic expansion in $\epsilon=4-d$ or directly in $d=3$ dimensions, the
corresponding corrections to the leading scaling behaviour automatically appear as a power series in the
scaling variable $(\lambda-\lambda^*)t^{\omega\nu}$.  These corrections have been studied in great detail,
for example in \cite{bagnuls85,bagnuls87,schloms}.  In the $1/N$ expansion studied here, this is not so.  After
the minimal renormalization that removes terms proportional to $\ln z$ from the correlation functions, there
remain weaker singularities of the form $z^n\ln z$, which will give rise to corrections of the form $x^n\ln x$.
Keeping only the leading term, proportional to $z\ln z$, the renormalized version of the constraint equation
(\ref{constraint}) is
\begin{equation}
t(z,\tilde{t})=\frac{1}{z+2a}\left[2a\tilde{t}^{1/2}\left(1+\frac{c'}{N}+\cdots\right)+z\tilde{t}\left(1+\frac{c}{N}\ln z
+\cdots\right)\right]\,,\label{appconstraint}
\end{equation}
where
\begin{equation}
c=\left(1-\frac{8a^2}{b^2}\right)\frac{S_3}{2b}=\frac{4}{\pi^2}\left(1-\frac{8}{\pi^2}\right)\,.
\end{equation}
and $c'$ is a number which must be evaluated numerically, though its exact value is immaterial for our present
purposes.  The ellipsis in (\ref{appconstraint}) indicate both terms of higher order in $N^{-1}$ and higher powers of $z$.
To obtain the scaling form of the inverse susceptibility $\tilde{t}$, we express the right-hand side of (\ref{appconstraint})
in the form of (\ref{rgsolution}) using the explicit characteristic functions (\ref{newcharacteristics}) and choose
$\ell=\tilde{t}^{1/(2-\eta)}$ to obtain
\begin{equation}
t=\frac{\tilde{t}^{1/\gamma}}{z+2a}\left[2a\left(1+\frac{c'}{N}+\cdots\right)+z\tilde{t}^{\omega\nu/\gamma}\left(1+\frac{c}{N}
\ln\left(z\tilde{t}^{\omega\nu/\gamma} \right)+\cdots \vphantom{\frac{c'}{N}}\right)\right]\,.
\end{equation}
On defining the scaling field $\tau$ and the scaling variable $y$ by
\begin{equation}
\tau=\left(\frac{1+z/2a}{1+c'/N+\cdots}\right)t\qquad\qquad y=\frac{z\tau^{\omega\nu}}{2a(1+c'/N+\cdots)}
\end{equation}
and the scaling function ${\cal T}(y)$ through $\tilde{t}=\tau^\gamma{\cal T}(y)$, we find that this scaling function is
the solution of
\begin{equation}
{\cal T}^{1/\gamma}\left\{1+y{\cal T}^{\omega\nu/\gamma}\left[1+\frac{c}{N}\ln y +\frac{c}{N}\ln\left(
2a{\cal T}^{\omega\nu/\gamma}\right)+\cdots\right]\right\}=1\,.
\end{equation}
For small $y$, it has the expansion
\begin{equation}
{\cal T}(y)=1-\gamma y\left(1+\frac{c}{N}\ln y\right)+\cdots\,.
\label{scalingwithlog}
\end{equation}
The presence of the term proportional to $y\ln y$ seems to be a novel result, which is discussed further in Section VI.

\section{Renormalization of the specific heat}

The specific heat exponent is given in three dimensions by
\begin{equation}
\alpha=2-3\nu=-1+\frac{16}{\pi^2N}+O(N^{-2})\,.
\end{equation}
As discovered long ago by Abe and Hikami \cite{abe},  the fact that $t^{-\alpha}=t+O(N^{-1})$ is associated with a kind
of degeneracy.  When exponents are obtained in the $1/N$ expansion by exponentiating logarithms of
$\bar{t}_0\equiv t_0-t_{0\rm{c}}$, it is essential to account correctly for a regular contribution proportional to $\bar{t}_0$.
Specifically, writing
\begin{equation}
C(\bar{t}_0)=C^{\vphantom{-\alpha}}_0\bar{t}_0^{-\alpha}+C_1\bar{t}_0\,,\label{formofc}
\end{equation}
it is argued in \cite{abe} that the coefficients $C_0$ and $C_1$, considered as functions of dimension $d$, have discontinuities
at $d=3$ that give rise to `anomalous' logarithms of $\bar{t}_0$ when the whole expression is expanded in powers of $1/N$.
Within the renormalization scheme proposed here, there are similar `anomalous' logarithms of $z$, which prevent the
specific heat from being multiplicatively renormalized.  To be concrete, we define
\begin{equation}
C(\lambda_0,\bar{t}_0)=N^{-1}\int d^3r\langle\phi^2(\bbox{r})\phi^2(\bbox{0})\rangle\,,
\end{equation}
where $\phi^2=\sum_i\phi_i^*\phi^{\vphantom{*}}_i$, and find that it is given to relative order $1/N$ by the sum of diagrams
shown in figure 5. We will define a renormalized function $C_{\mathrm{R}}(z,\tilde{t})$ by
\begin{equation}
C(\lambda_0,\bar{t}_0)=C(\lambda_0,0)+C_1(\lambda_0)\bar{t}_0 +\mu^{-1}\bar{Z}_t^{-2}(z)C_{\mathrm{R}}(z,\tilde{t})\,,
\label{crenorm}
\end{equation}
where $\bar{Z}_t(z)=(1+2a/z)Z_t(z)$ is the same renormalization factor that appears in (\ref{trenorm}). Dimensional analysis
tells us that $C_1(\lambda_0)=C_1\lambda_0^{-3}$, where $C_1$ is a number.  Were we able to supply it, an analysis
to all orders in $1/N$ of the divergences of the theory as $z\to 0$ would tell us whether an appropriate choice of this
number (at each order in $1/N$) is sufficient to make $C_{\mathrm R}(0,\tilde{t})$ finite.  This is beyond our present skill,
but we have verified that the procedure does work (with $C_1=-4/b^2$) at next-to leading order.  In perturbative
renormalization schemes, as is well known, an additive renormalization is also required to make a quantity
analogous to $C_{\mathrm R}$ finite in the limit $d\to 4$.  In that case, the additive term depends on the renormalization
scale $\mu$ and, in consequence, the renormalized specific heat obeys an inhomogeneous renormalization-group
equation (see, for example \cite{brezin}).  Here, by contrast, the first two terms on the right-hand side of
(\ref{crenorm}) are functions only of $\lambda_0$ and $t_0$, so $C_{\mathrm R}$ obeys a homogeneous equation of
the form (\ref{rgeqn}).

Finally, let us write (\ref{crenorm}) in the form
\begin{equation}
\bar{Z}_t^2\left[C(\lambda_0,\bar{t}_0)-C(\lambda_0,0)\right] =\bar{C}_0t^{-\alpha}+\bar{C}_1t\,,
\end{equation}
with $\bar{C}_0=C_{\mathrm R}(0,1)$ and $\bar{C}_1=C_1\bar{Z}_t^3\lambda_0^{-3}$.  It is somewhat reassuring to
find that the amplitude ratio can be written as
\begin{equation}
\frac{\bar{C}_0}{\bar{C}_1}=\frac{\pi^2}{4}-1+{\rm O}(N^{-1})
\end{equation}
in agreement with that given in \cite{abe}, though $\bar{C}_1$ does not, in general, have a well-defined
limit as $z\to 0$.

\section{Discussion}

We have proposed a renormalization scheme within which critical exponents and scaling functions for an
$N$-vector Ginzburg-Landau model can be estimated by means of a systematic expansion in $1/N$.  From
the field-theoretic point of view, the essential function of the renormalization group is to relate the values of
correlation functions in the critical region, where infrared singularities arise from the vanishing `mass'
$t\sim T-T_{\mathrm c}$, to their values when $t$ is of order 1, as exhibited in (\ref{rgsolution}). To ensure that
critical singularities are correctly exponentiated, a renormalization scheme must ensure that these singularities
do not reappear from the running of other parameters.  In the context of the $1/N$ expansion, this potentially
happens when the inverse quartic coupling constant $z\sim\lambda^{-1}$ vanishes, and the essential feature of
our renormalization scheme is to ensure that renormalized correlation functions have finite limits when
$z\to 0$. In 3 dimensions, the minimal way of achieving this is to subtract leading powers of $\ln z$, and this
procedure does indeed serve to recover the standard results for critical exponents.

The main purpose of this paper is to explain how the scheme works, in preparation for a detailed investigation of the
critical properties of high-temperature superconductors to be described in \cite{lee}.  However, in the course of
the formal investigation reported here, we have encountered a general feature that seems to have been
unsuspected hitherto.  It has usually been taken for granted that the approach to a critical point can be described
by expressing the Hamiltonian in the form ${\cal H}={\cal H}^*+\sum_ig_i{\cal O}_i$, where ${\cal H}^*$ is an
infrared stable fixed-point Hamiltonian and the ${\cal O}_i$ are eigenoperators of the renormalization group
at this fixed point (see, for example \cite{fisher74,fisher98}).  By expanding correlation functions in powers of the
coefficients $g_i$, one expects to obtain corrections to the asymptotic critical singularities in the form of
powers of the scaling variables $g_i\tau^{\Delta_i}$, with exponents $\Delta_i$ determined by the eigenvalues
of the ${\cal O}_i$.  In treatments based on perturbative expansions in the coupling constant $\lambda$, this
expectation is realized automatically in the case of corrections of the form $(\lambda-\lambda^*)\tau^{\omega\nu}$
associated with departures of $\lambda$ from its fixed-point value $\lambda^*$.  Within the non-perturbative
approximation scheme afforded by the $1/N$ expansion, however, this does not happen.  We see explicitly
that the fixed point corresponds to $z=\lambda^{-1}=0$, and that correlation functions do not possess power-series
expansions in $z$.  Although scaling is maintained, in the sense that corrections appear in terms of the scaling
variable $y\propto z\tau^{\omega\nu}$, scaling functions such as that exhibited in (\ref{scalingwithlog}) for the
inverse susceptibility are not expressible as power series in $y$.

At relative order $1/N$, the non-analyticity of the scaling functions is logarithmic, and this is probably true at
higher orders also.  These are not, however, the same as the well-known logarithmic corrections that occur at
the upper critical dimension $d=4$ \cite{larkin,wegner,brezin} where $\phi^4$ is a marginal operator.  The latter
arise from a degeneracy in the renormalization-group equations which destroy the scaling property, yielding
logarithms of $\tau$ rather than of a scaling variable such as $y$. In fact, our renormalization scheme is restricted
to $d=3$;  we do not know in detail how to formulate a similar scheme in general dimensions.  Quite possibly,
the singularities we encountered appear, like the Abe-Hikami specific heat anomaly, only at special, rational
values of $d$, and it is not clear whether they need be logarithmic in general.  It is also far from clear whether
they are of more than academic interest.  In principle, they should presumably be present in, for example, the
specific heat of $^4{\rm He}$ near the lambda transition (corresponding to $N=1$), where rather precise
measurements of critical properties have been made \cite{lipa83,lipa96}.  The data are reasonably consistent with
the assumed power-law correction, but may well not be precise enough to detect a logarithmic factor. Unfortunately,
while the non-perturbative nature of the $1/N$ expansion is helpful in indicating the presence of these logarithms,
its notoriously poor convergence make it hard to estimate the likely sizes of the coefficients that multiply them.

\section*{Acknowledgments}

The financial support of the Engineering and Physical Sciences Research Council through Research Grant
GR/M09261 is gratefully acknowledged.

\appendix

\section{Singular parts of Feynman integrals}
To implement our renormalization scheme at next-to-leading order in the $1/N$ expansion, we need to know
the singularities of the integrals (\ref{aintegral}) and (\ref{b1integral}) - (\ref{b3integral}) near $z= 0$.  Each of them
can be expressed in terms of integrals of the form
\begin{equation}
I_K=\int\frac{d^3k}{(2\pi)^3}\left[z+\frac{b}{k}-\frac{m}{k^2}\tilde{\Pi}\left(\frac{m^2}{k^2}\right)\right]^{-1}\frac{1}{k^4}
K\left(\frac{m^2}{k^2}\right)\,,
\end{equation}
or $dI_K/dz$, where $k=\vert\bbox{k}\vert$ and $K(m^2/k^2)$ is proportional to $k^4$ as $k\to 0$, but approaches
a finite value as $k\to\infty$.  The singularities arise from the region of integration where $k$ is large. To evaluate
them, we write
\begin{equation}
I_K=S_3\int_1^\infty\frac{dk}{k(zk+b)}\left[1-\frac{(m/k)\tilde{\Pi}(m^2/k^2)}{zk+b}\right]^{-1}K\left(\frac{m^2}{k^2}\right)
+\mathrm{reg}\,,
\end{equation}
where $S_3=1/2\pi^2$ and `reg' denotes contributions that are regular when $z\to 0$, arising here from the integration
region $0\le k\le 1$. Defining
\begin{equation}
f^{(\ell\,)}\!\left(\frac{m}{k}\right)=\left[ \left(\frac{m}{k}\right)^2 \tilde{\Pi}\left(\frac{m^2}{k^2}\right)\right]^\ell K\left(\frac{m^2}{k^2}\right)\,,
\end{equation}
and using the series expansion $f^{(\ell)}(x)=\sum_nf^{(\ell)}_n(x)$, we have
\begin{equation}
I_K=\sum_{\ell,n}f^{(\ell)}_n\,I_{\ell,n} + \mathrm{reg}\,,
\end{equation}
where
\begin{eqnarray}
I_{\ell\,,n}&=&S_3\int_1^\infty\frac{dk}{k(zk+b)^{\ell\,+1}}\left(\frac{m}{k}\right)^{n-\ell}
=\frac{1}{\ell\,!}\left(-\frac{1}{m}\frac{\partial}{\partial z}\right)^\ell I_{n}\,.\\
I_n&=& S_3\int_1^\infty\frac{dk}{k(zk+b)}\left(\frac{m}{k}\right)^{n}\,.
\end{eqnarray}
The integral $I_0$ is
\begin{equation}
I_0=S_3\int_1^{\infty}\frac{dk}{k(zk+b)} =-\frac{S_3}{b}\ln z +\mathrm{reg}
\end{equation}
and for $n\ge 1$, the recursion relation
\begin{equation}
I_n=\frac{S_3m}{b}\int_1^\infty\frac{dk}{k}\left(\frac{1}{k}-\frac{z}{zk+b}\right)\left(\frac{m}{k}\right)^{n-1}
=\frac{S_3m^n}{nb}-\frac{mz}{b}I_{n-1}
\end{equation}
is readily solved to yield
\begin{equation}
I_n=\frac{S_3m^n}{b}\sum_{j =\, 0}^{n-1}\frac{(-1)^j}{(n-j)}\left(\frac{z}{b}\right)^{n-j}+ \left(-\frac{mz}{b}\right)^nI_0
=-\frac{S_3}{b}\left(-\frac{mz}{b}\right)^n \ln z+ \mathrm{reg}
\end{equation}

With these results in hand, we can evaluate $I_K$ as
\begin{eqnarray}
I_K&=&-\frac{S_3}{b}\sum_{\ell,\,n}\frac{1}{\ell\,!}f^{(\ell)}_n\left(-\frac{1}{m}\frac{\partial}{\partial z}\right)^\ell
\left[\left(-\frac{mz}{b}\right)^n\ln z\right] + \mathrm{reg}\nonumber\\
&=&-\frac{S_3}{b}\ln z\sum_\ell\frac{1}{\ell\,!}\left(-\frac{1}{m}\frac{\partial}{\partial z}\right)^\ell f^{(\ell)}\!\left(-\frac{mz}{b}\right)
+ \mathrm{reg} \nonumber\\
&=&-\frac{S_3}{b}\ln z\sum_\ell\frac{1}{\ell\,!}\left(-\frac{1}{m}\frac{\partial}{\partial z}\right)^\ell\left\{\left[\left(\frac{mz}{b}\right)^2
\tilde{\Pi}\left(\frac{m^2z^2}{b^2}\right)\right]^\ell K\!\left(\frac{m^2z^2}{b^2}\right)\right\}+ \mathrm{reg}\nonumber\\
&=& -\frac{S_3}{b}\ln z\int_{-\infty}^\infty dq\,K(q^2)\sum_\ell \frac{1}{\ell\,!}\left[\frac{q^2\Pi(q^2)}{b}\right]^\ell
\left(\frac{\partial}{\partial q}\right)^\ell \delta\left(q-\frac{mz}{b}\right) + \mathrm{reg}\nonumber\\
&=& -\frac{S_3}{b}\ln z\int_{-\infty}^\infty dq\,K(q^2)\,\delta\left(q-\frac{mz}{b}+\frac{1}{b}q^2\tilde{\Pi}(q^2)\right)
+ \mathrm{reg}\,.
\end{eqnarray}
The factor of $\ln z$ can be extracted because $(\partial/\partial z)^{\ell}(z^p\ln z)=\ln z(\partial^\ell z^p/\partial z^\ell)+\mathrm{reg}$
when $p\ge 2\ell$.  Carrying out the $q$ integral, we finally obtain
\begin{equation}
I_K^{\mathrm{sing}}=-\frac{S_3\ln z}{b}\,K(Q^2)\left\{1+\frac{1}{b}\frac{d}{d Q}\left[Q^2\tilde{\Pi}(Q^2)
\right]\right\}^{-1}\,,
\end{equation}
where $Q(z)$ is the solution of
\begin{equation}
bQ+Q^2\tilde{\Pi}(Q^2)=mz\,.\label{qeqn}
\end{equation}
Solving for $Q$ as a power series in $z$, we obtain
\begin{eqnarray}
Q(z)&=&\frac{mz}{b}\left[1-\frac{4a}{b^2}\,mz+\frac{32a^2}{b^4}\,(mz^2)+O\left((mz)^3\right)\right]\\
I_K^{\mathrm{sing}}&=&-\frac{S_3}{b}\ln z\left[1-\frac{8a}{b^2}\,mz +\frac{96a^2}{b^4}\,(mz)^2+O\left((mz)^3\right)\right]K(Q^2)\,.
\end{eqnarray}

%%%%%%%%%%%%%%%%%%%%%%%%%%%%%%%%%%%%%%%%
%%%%%%%%%%%%%%%%%%%%%%%%%%%%%%%%%%%%%%%%

\begin{figure}
\caption{Elements of Feynman diagrams in the $1/N$ expansion:  ({\it a}) the $\psi$ propagator; ({\it b}) the counterterm $f(\tilde{t}_0,\lambda_0)$;
({\it c}) the first few terms in the expansion of the propagator $\Delta(\bbox{r},\bbox{r}';\Psi)$; ({\it d}) elementary vertex functions arising from the
expansion of $NH_{\mathrm{eff}}$.  Solid lines represent the lowest-order $\phi$ propagator $\Delta(\bbox{r}-\bbox{r}')$ and filled circles represent the
counterterm $\delta$.  In ({\it c}) and ({\it d}), the dashed lines are the legs to which $\psi$ propagators can be attached; arbitrary sequences of these
lines and the dashed legs are allowed.}
\label{fig1}
\end{figure}
\begin{figure}
\caption{Diagrammatic representation of the constraint equation at next-to-leading order.}
\label{fig2}
\end{figure}
\begin{figure}
\caption{Diagrammatic representation of the order-parameter 2-point function at next-to-leading order.}
\label{fig3}
\end{figure}
\begin{figure}
\caption{Diagrammatic representation of the vertex function $\Gamma^{(4)}$ at next-to-leading order.}
\label{fig4}
\end{figure}
\begin{figure}
\caption{Diagrammatic representation of the specific heat at next-to-leading order.  In the first diagram, the cross indicates that the loop contains
two $\phi$ propagators.}
\label{fig5}
\end{figure}

\end{document}